\documentclass{pasj01}
\Received{$\langle$reception date$\rangle$}
\Accepted{$\langle$acception date$\rangle$}
\Published{$\langle$publication date$\rangle$}
\usepackage{color}

\usepackage{deluxetable}
\makeatletter
\def\@to{to}
\makeatother
\usepackage{multirow}
\usepackage{wrapfig}

\begin{document}

\title{The formation of the young massive cluster B1 in the Antennae galaxies (NGC 4038/NGC 4039) triggered by cloud-cloud collision}
\author{Kisetsu Tsuge\altaffilmark{1}, Kengo Tachihara\altaffilmark{1}, Yasuo Fukui\altaffilmark{1, 2}, Hidetoshi Sano\altaffilmark{3}, Kazuki Tokuda\altaffilmark{4,5}, Junko Ueda\altaffilmark{3}, Daisuke Iono\altaffilmark{3,6}}%
\altaffiltext{1}{Department of Physics, Nagoya University, Furo-cho, Chikusa-ku, Nagoya 464-8601, Japan}
\altaffiltext{2}{Institute for Advanced Research, Nagoya University, Furo-cho, Chikusa-ku, Nagoya 464-8601, Japan}
\altaffiltext{3}{National Astronomical Observatory of Japan, 2-21-1 Osawa, Mitaka,Tokyo, 181-8588, Japan}
\altaffiltext{4}{Department of Physical Science, Graduate School of Science, Osaka Prefecture University, 1-1 Gakuen-cho, Naka-ku, Sakai, Osaka 599-8531, Japan}
\altaffiltext{5}{Chile Observatory, National Astronomical Observatory of Japan, National Institutes of Natural Science, 2-21-1 Osawa, Mitaka, Tokyo 181-8588, Japan}

\altaffiltext{6}{SOKENDAI (The Graduate University for Advanced Studies), 2-21-1 Osawa, Mitaka, Tokyo 181-8588, Japan}
\email{tsuge@a.phys.nagoya-u.ac.jp}

\maketitle

\begin{abstract}
The Antennae Galaxies is one of the starbursts in major mergers. Tsuge et al. (2020) showed that the five giant molecular complexes in the Antennae Galaxies have signatures of cloud-cloud collisions based on the ALMA archival data at 60 pc resolution. In the present work we analyzed the new CO data toward the super star cluster (SSC) B1 at 14 pc resolution obtained with ALMA, and confirmed that two clouds show complementary distribution with a displacement of $\sim$70 pc as well as the connecting bridge features between them. The complementary distribution shows a good correspondence with the theoretical collision model (Takahira et al. 2014), and indicates that {the formation of SSC B1 having $\sim$10$^{6}$ $M_{\rm \odot}$ was consistent with the trigger of cloud-cloud collision} with a time scale of $\sim$1Myr, which is consistent with the cluster age. It is likely that SSC B1 was formed from molecular gas of $\sim$10$^7$ $M_{\rm \odot}$ with a star formation efficiency of $\sim$10 \% in 1 Myr. We identified a few places where additional clusters are forming. Detailed gas motion indicates stellar feedback in accelerating gas is not effective, while ionization plays a role in evacuating the gas around the clusters at a $\sim$20-pc radius. The results have revealed the details of the parent gas where a cluster having mass similar to a globular is being formed.
\end{abstract}

\section{Introduction}
Super star clusters (SSCs) are attracting keen interest of the researchers, since SSCs are highly energetic and affect the galactic evolution substantially. SSCs are also supposed to be a present-day analog to the ancient globulars, whose formation is deeply connected to the environment of the early universe. The Antennae Galaxies is an outstanding major merger where {thousand}s of young SSCs are discovered.

\begin{figure*}[htbp]
\begin{center}
\includegraphics[width=\linewidth]{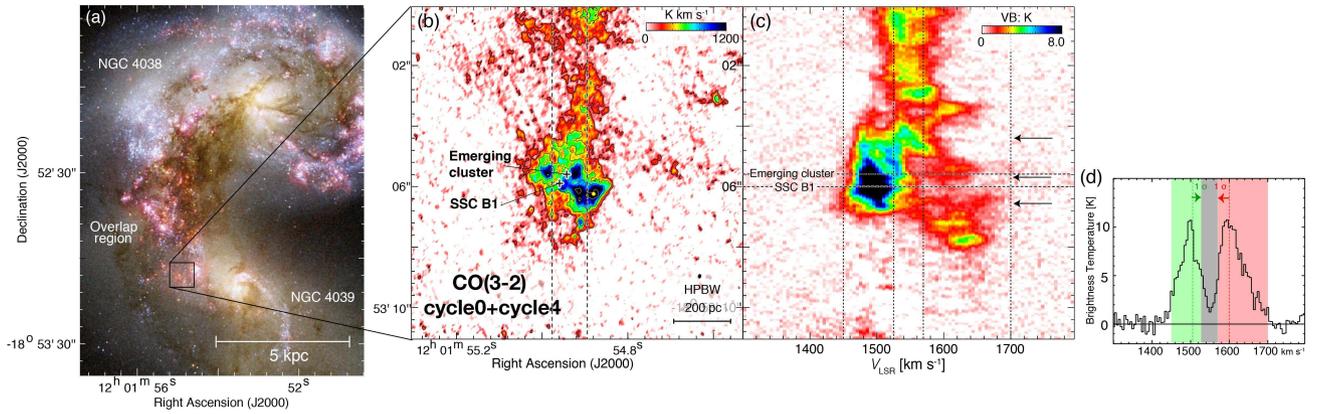}
\end{center}
\caption{ (a) An optical image of NGC 4038/4039 produced by the Hubble Space Telescope data. B-band image is shown in blue, V-band image in green, and a combination of the I-band and H$\alpha$ images in red. (b) Total integrated intensity map of $^{12}$CO (3--2) toward superstar cluster B1. The black {crosses show the positions of SSC B1 and {an} emerging cluster classified by radio continuum, CO, and optical/{near}-infrared data (Whitmore et al. 2014).} The lowest contour level and intervals are 250 and 250 K km s$^{-1}${, respectively}. The dashed lines show the integration range in R. A. in (c). (c) Declination--velocity diagram of $^{12}$CO(3--2). The black arrows indicate the position of bridge features. {The horizontal dashed line{s} show the positions of SSC B1 {and the emerging cluster.}} (d) Typical spectrum of $^{12}$CO($J$=3--2) toward integrated intensity peak at the position of yellow circle (R.A., Dec.)=( $\timeform{12h01m54.88s}$, $\timeform{-18D53'6.24"}$). The horizontal axis and vertical axis indicate $V_\mathrm{LSR}$ [km s$^{-1}$] and intensity [K], respectively. Green and red shaded regions are $V$$_{\rm LSR}$=1450 to 1525 km s$^{-1}$ (blue-shifted cloud) and 1550 to 1700 km s$^{-1}$ (redshifted cloud), respectively. The green and red dashed vertical lines show the average velocity of blue shifted and red shifted clouds, respectively. Grey shaded region shows velocity range of bridge features in the intermediate velocity range between the blue-shifted and red-shifted clouds. The velocity ranges of shaded area correspond to the integration ranges as shown in Figure 2.}  
\label{fig1}
\end{figure*}%

Genzel et al.(1998) showed that the galaxy mergers have excess infrared luminosity as compared with the isolated galaxies, lending support for the active star formation triggered by galaxy interactions in mergers. The Antennae Galaxies is the most outstanding major merger closest to the Milky Way at a distance of 22 Mpc (Schweizer et al. 2008). The galaxies consist of the two spirals NGC~4038 (R.A. = $\timeform{12h01m53.00s}$, Dec. = $\timeform{-18D52'10"}$) and NGC~4039 (RA=$\timeform{12h01m53.60s}$, Dec=$\timeform{-18D53'11"}$). The interaction between NGC~4038 and NGC~4039 has been continuing probably since a few 100 Myr ago (Mihos et al. 1993; Karl et al. 2010; Renaud et al. 2015 and references therein). The Antennae is prominent because of the unusually active star formation {(often called ''starbursts'')} including {thousands} of young {massive} SSCs (e.g., Whitmore \& Schweizer 1995), and clear signatures of the galactic interaction as the two long tails.

In a smaller distance, we have increasing evidence for similar cloud-cloud interactions triggering the formation of SSCs, which include R136 in the Large Magellanic Cloud (Fukui et al. 2017) and NGC604 in M33 (Tachihara et al. 2018) as well as several SSCs in the Milky Way (Furukawa et al. 2009; Ohama et al. 2010; Fukui et al. 2014; 2016; Kuwahara et al. 2020; Fujita et al. 2020).

Molecular observations are crucial in studying cluster formation in mergers. In particular, ALMA offers an ideal opportunity to investigate the molecular gas distribution and kinematics at the highest resolution achieved so far, which are possibly related not only to the feedback by the SSCs but also directly to the SSC formation. 

\begin{figure*}[htbp]
\begin{center}
\includegraphics[width=16cm]{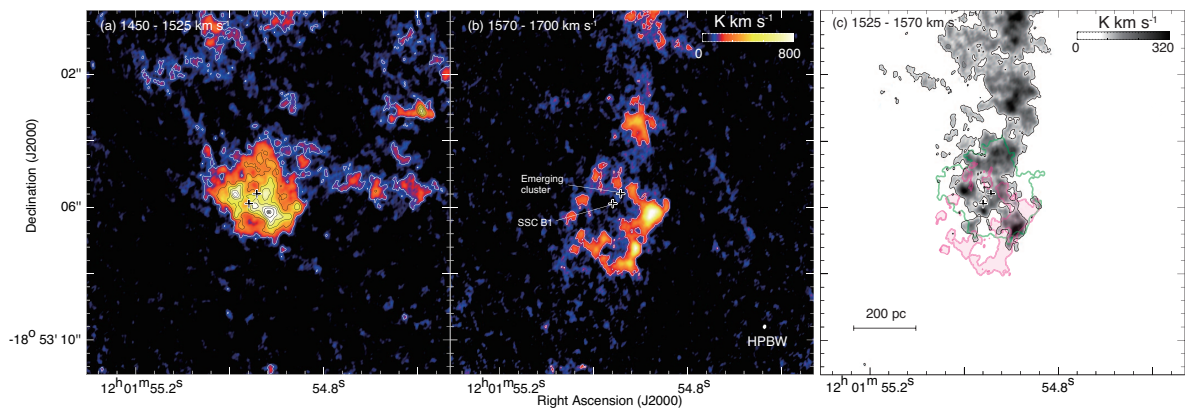}
\end{center}
\caption{(a) The spatial distribution of blue{-}shifted cloud. The integration velocity range of $V$$_{\rm LSR}$ = 1450--1525 km s$^{-1}$. The contour levels are 300, 450, 600,750, 900,1060, 1070, and 1090 K km s$^{-1}$. (b) The spatial distribution of red{-}shifted cloud. The integration velocity range of $V$$_{\rm LSR}$ = 1570--1700 km s$^{-1}$. (c) Intensity map of $^{12}$CO($J$=3--2) consisting of three velocity components. The black image indicates bridge features ($V$$_{\rm LSR}$ = 1525--1570 km s$^{-1}$). The red and green contours indicate the red-shifted cloud and blue-shifted cloud, respectively. The contour levels of three velocity components are 40 $\sigma$ (95 K km s$^{-1}$ for bridge; 156 K km s$^{-1}$ for red-shifted cloud; 120 K km s$^{-1}$ for blue-shifted cloud). {The symbols are the same as in Figure 1.}} 
\label{fig2}
\end{figure*}%

The previous works on the ALMA data investigated the distribution of molecular clouds and dense gas (Whitmore et al. 2014; Schirm et al. 2016), the effect and mechanism of stellar feedback (Herrera et al. 2017), and progenitors of SSCs (Herrera et al. 2011, 2012; Johnson et al. 2015), whereas the mechanism of cluster formation has not been well understood. Whitmore et al. (2014) briefly suggested a possible role of filamentary cloud collision induced by tidal interaction in SSC formation, but no further exploration of cluster formation was undertaken in these articles. Most recently, Tsuge et al. (2020) {(hereafter Paper I)} studied the ALMA CO data and presented evidence for cloud-cloud collisions in the five giant molecular cloud complexes (SGMCs), opening a new window in the pursuit of SSC formation. These authors used ALMA Cycle0 data at 50 pc resolution, while the Firecracker, a SSC candidate, was studied at 10 pc resolution of Cycle 4 (Finn et al. 2019) and complementary distribution between two colliding clouds was revealed.

\begin{figure*}[htbp]
\begin{center}
\includegraphics[width=\linewidth]{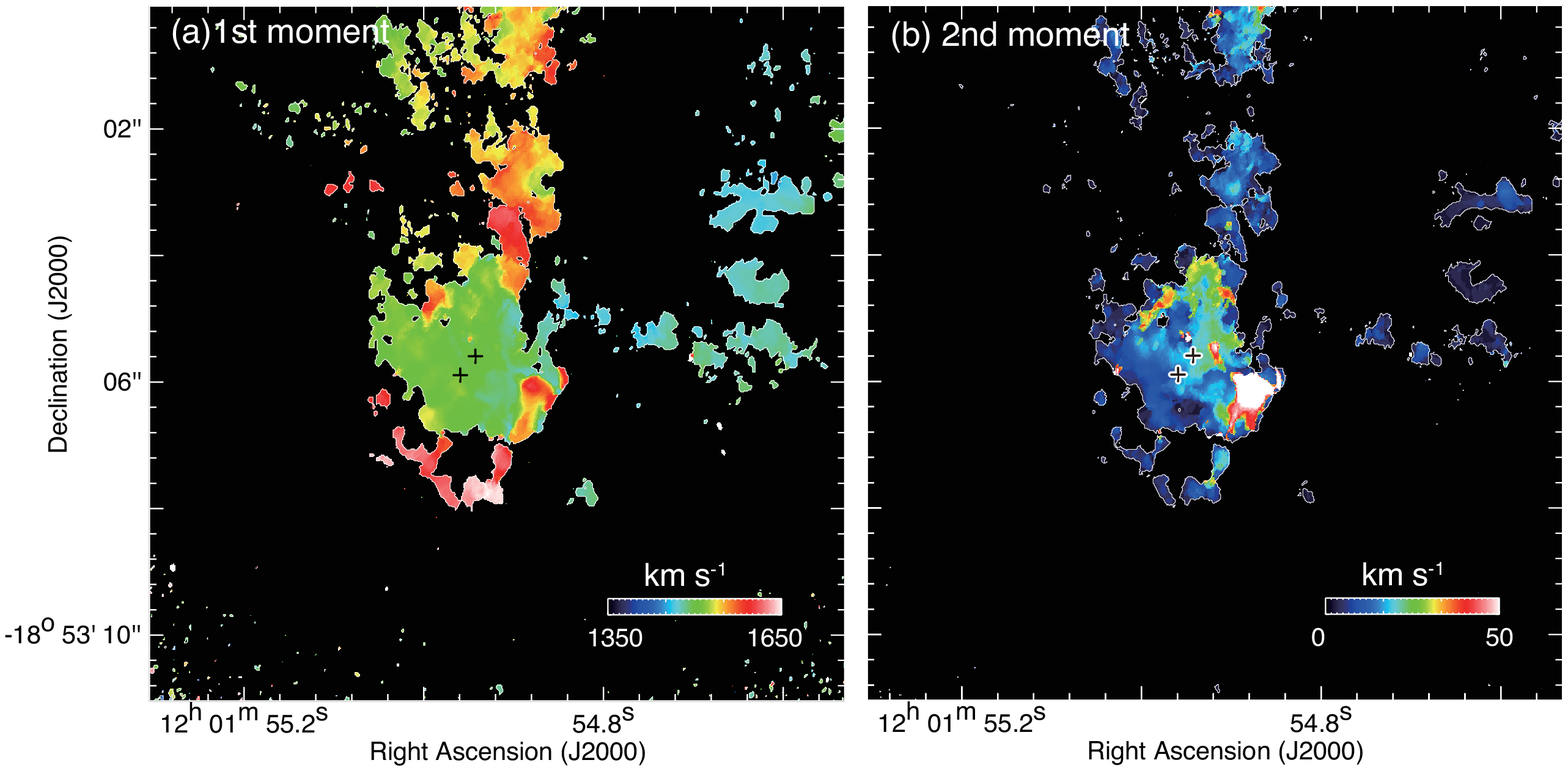}
\end{center}
\caption{ {(a) 1st moment map of $^{12}$CO(3--2) toward SSC B1. A velocity range $V$$_{\rm LSR}$= 1350--1650 km s$^{-1}$ is used for calculation. (b) 2nd moment map of $^{12}$CO(3--2) toward SSC B1. The same velocity range is used for calculation.} {The symbols are the same as in Figure 1.}} 
\label{fig3}
\end{figure*}%

In the Antennae, 40 \% of youngest SSCs are located in the overlap region as indicated in Figure 1a, where two galaxies are merging (Wilson et al. 2000). The SSCs are cataloged by using the near infrared data taken with Hubble Space Telescope (Gilbert \& Graham 2007; Whitmore et al. 2010). {$\sim$60\% of the 50 most massive clusters whose mass is greater than $\sim$6$\times$10$^5$ $M_{\odot}$ 
are located in the overlap region (Whitmore et al. 2010).  }
{Five} out of 8 SSCs whose mass is larger than 10$^{6}$ $M_{\rm \odot}$ and age is younger than 10 Myr are concentrated to the southern part of the overlap region as shown in Figure 1b. We first focus on SSC B1, the most luminous embedded cluster candidate in the Antennae, which was identified by Whitmore \& Schweizer (1995) as WS 80 (IR source). This cluster received much attention as the strongest CO source (Wilson et al. 2000), the strongest ISO source (Vigroux et al. 1996; Mirabel et al. 1998), and the strongest radio source (Neff \& Ulvestad 2000). Wilson et al. (2000) briefly suggested a possibility of collision between super giant molecular complexes SGMCs as the origin of strong mid-infrared emission, whereas more details of the SSC formation was not discussed. 

\begin{figure*}[htbp]
\begin{center}
\includegraphics[width=\linewidth]{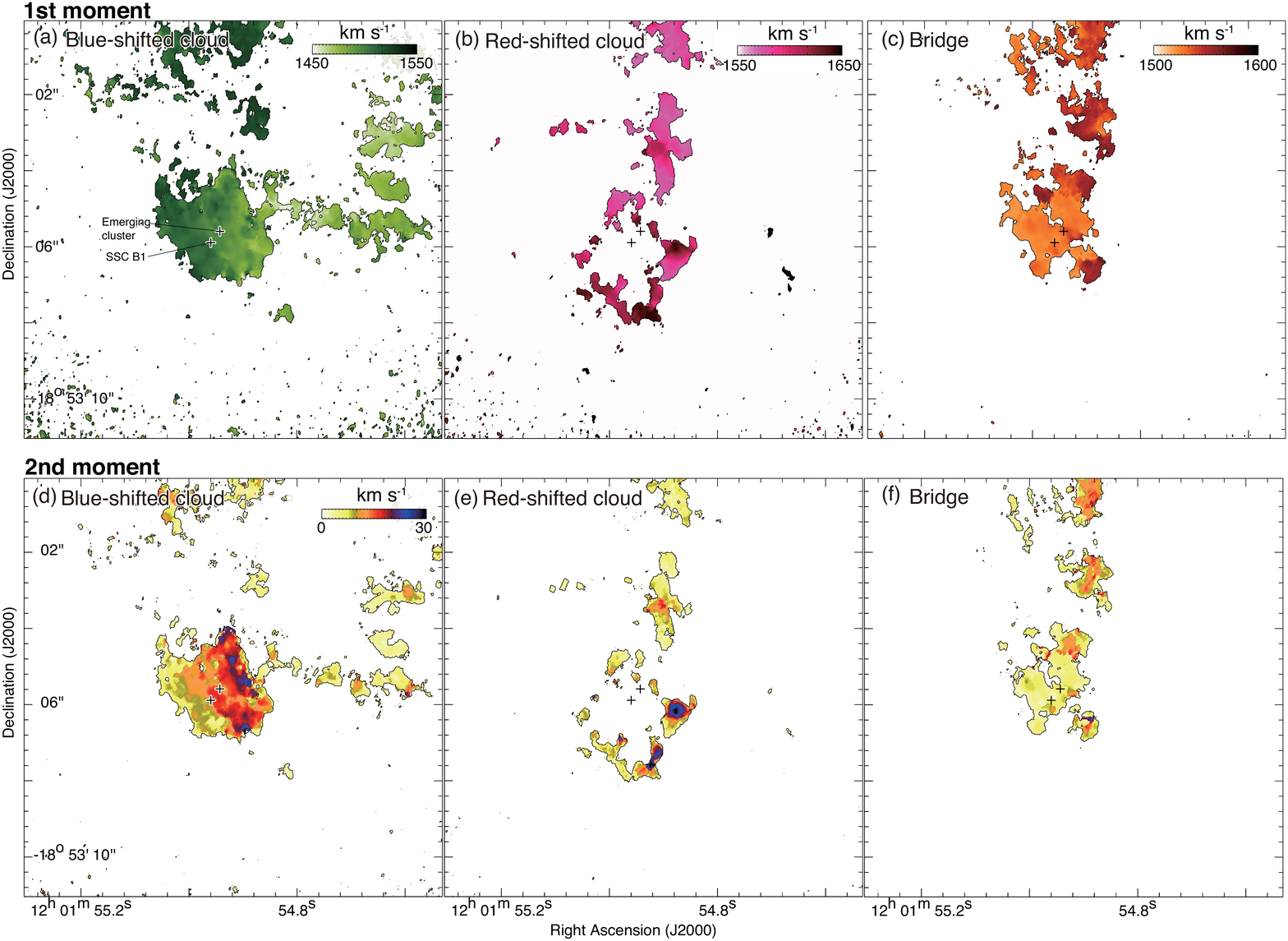}
\end{center}
\caption{ {(a--c) 1st moment map of $^{12}$CO(3--2) toward SSC B1. Velocity ranges {used} for calculating momentum are $V$$_{\rm LSR}$= 1450--1525 km s$^{-1}$ (blue-shifted cloud) for (a), $V$$_{\rm LSR}$ = 1570--1700 km s$^{-1}$ (red-shifted cloud) for (b), and {$V$$_{\rm LSR}$ = 1525--1570 km s$^{-1}$ (bridge) for (c)}. {(d--f)} 2nd moment map of $^{12}$CO(3--2) toward {SSC B1}. Velocity ranges {used} for calculating moment are the same as in (a), (b), and (c). {The symbols are the same as in Figure 1.}}} 
\label{fig4}
\end{figure*}%

In the present paper, we analyze the ALMA Cycle3 data of CO 3--2 toward SSC B1 and present detailed gas distribution associated with the cluster. Section 2 described the datasets, and Section 3 the results of the analysis. We discuss the implications in SSC formation and feedback in Section 4 and give conclusions in Section 5.

\section{The ALMA archive of the Antennae and data reduction}
In the present work we made use of the archival {$^{12}$}CO($J$=3--2) data of Band 7 (345 GHz) of the Antennae (NGC 4038/39) from Cycle 0 project 2011.0.00876 (P.I., B. Whitmore). The detailed descriptions of the observations are given by Whitmore et al. (2014). We re-did the imaging process from the visibility data by using the CASA (Common Astronomy Software Application) package (McMullin et al. 2007) version 5.0.0. We used the {tclean task with} multiscale CLEAN algorithm (Cornwell 2008) implemented in the CASA. The resultant synthesized beam size is 0$\farcs$70 $\times$ 0$\farcs$46 and the 1$\sigma$ RMS noise is $\sim$4.8$\times$10$^{-3}$ Jy/beam at a velocity resolution of 5.0 km s$^{-1}$, which corresponds to a surface brightness sensitivity of $\sim$0.15 K.

We also reduced the higher resolution $^{12}$CO($J$=3--2) data of Cycle 3 project 2015.1.00038S (P.I., C. Herrera) toward SSC B1. {ALMA Cycle 3 Band 7 observations have been carried out in September 2016 using 36 antennas of the 12-m main arrays. The observations used the single-pointing mode centered at ($\alpha$J2000, $\delta$J2000)$\sim$($\timeform{12h01m54.9s}$, $\timeform{-18D53'4.0"}$). The baseline length ranges from 27.09 to 3143.76 m, corresponding to $u$-$v$ distances from 31.25 to 3626.08 k$\lambda$. Three quasars, J1256-0547, J1203-1612, and J1215-1731 were observed as the complex gain calibrator, the flux calibrator, and the phase calibrator, respectively. To reduce the datasets, we also used CASA package version 5.5.0 and multiscale CLEAN algorithm with the natural weighting. We finally combined the individually imaged Cycle 0 and Cycle 3 data sets using the feathering task implemented in CASA. The combined baseline length ranges from 21.36 to 3143.76 m, corresponding to $u$-$v$ distances from 24.64 to 3626.08 k$\lambda$. Here, the maximum recoverable scale (MRS) is to be 2$\farcs$ 64.}
The resultant synthesized beam size is $\sim$ 0$\farcs$15$\times$0$\farcs$11 {for natural weighting}, which is four times better than Cycle 0 data. {The typical spatial resolution is $\sim$14 pc at the distance of the Antennae Galaxies.} The 1$\sigma$ RMS noise is $\sim$1.0$\times$10$^{-2}$ Jy/beam at a velocity resolution of 5.0 km s$^{-1}$, which corresponds to a surface brightness sensitivity of $\sim$0.6 K. In order to investigate the missing flux, we compared the peak brightness temperature of the single-dish (JCMT) data (Zhu et al. 2003) and our reprocessed data smoothed to the single-dish resolution, $\timeform{14"}$. Although the ALMA flux is $\sim$20\% smaller than the single-dish data, the missing flux is considered to be not significant{, because larger structures than MRS are not discussed in this paper}.

\begin{figure*}[htbp]
\begin{center}
\includegraphics[width=\linewidth]{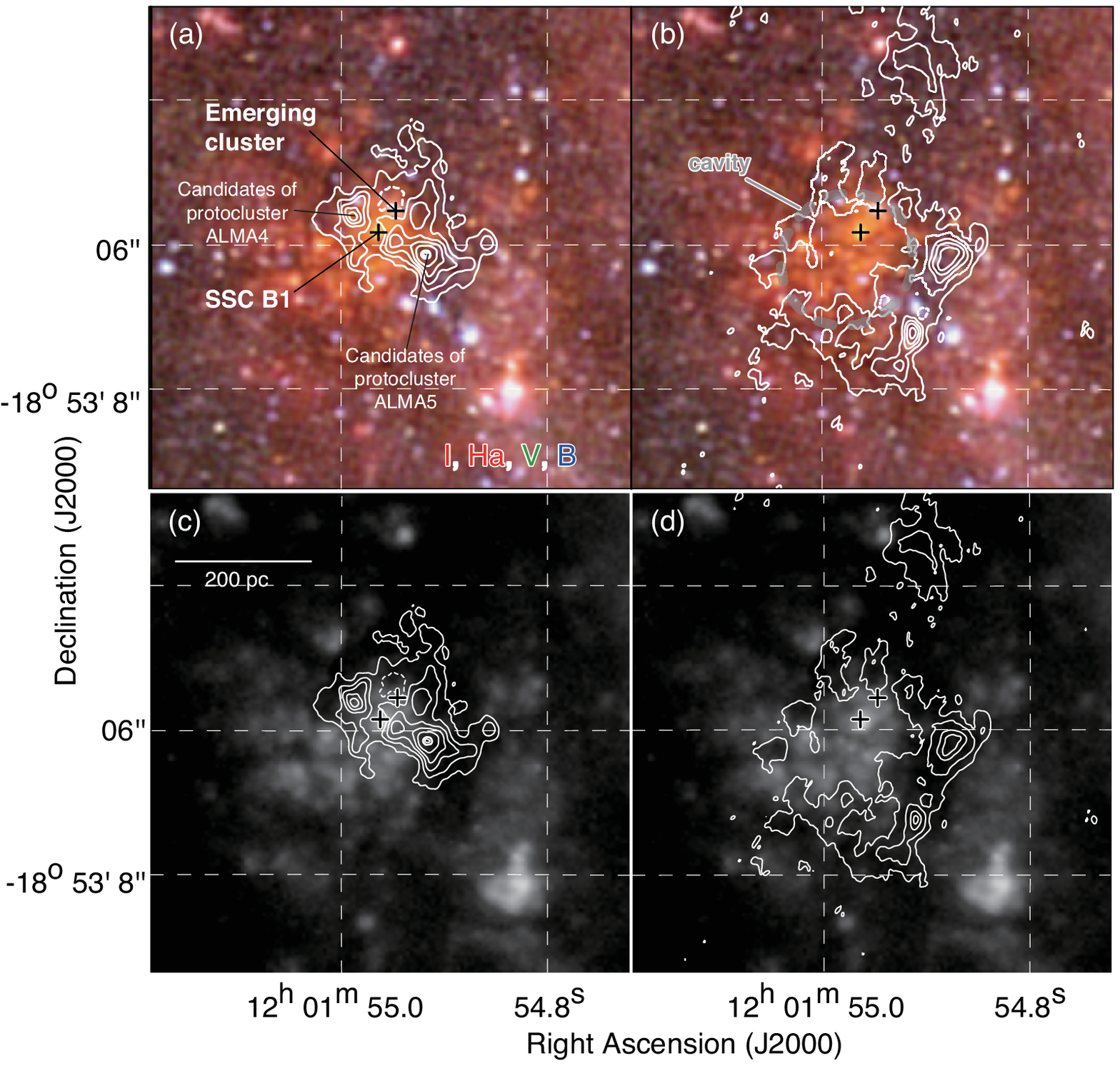}
\end{center}
\caption{(a)(b) False color image toward SGMC 4/5 obtained with HST overlaid with integrated intensity distribution of$^{12}$CO($J$=3--2) for the blue-shifted cloud and red-shifted cloud {by contours}. B-band image is shown in blue, V-band image in green, and a combination of the I-band and H$\alpha$ images in red.  The integrated velocity ranges are the same as in Figure 2. The contour levels are 300, 450, 600, 750, 900,1060, 1070, and 1090 K km s$^{-1}$ for (a); 120, 220, 320, 420, 520, 620, 720, 820, 920, 1020 K km s$^{-1}$ for (b). (c)(d) Integrated intensity distribution of $^{12}$CO($J$=3--2) for the blue-shifted cloud and red-shifted cloud by contours superposed on H$\alpha$ image obtained with $HST$. The integrated velocity ranges and contour levels of (c) and (d) are the same as in (a) and (b). {The symbols are the same as in Figure 1.} } 
\label{fig4-2}
\end{figure*}%


\section{Results}
\subsection{CO Distribution}

Figure 1a shows the optical image of the Antennae Galaxies and Figure 1b shows the distribution of the $^{12}$CO ($J$=3--2) emission obtained with ALMA Cycle 0 in the overlap region combined with Cycle 3 data. Spatial resolution is improved by $\sim$4 times, and cluster scale distribution is resolved at $\sim$10 pc with a peak toward SSC B1. 
{SSC B1 and the emerging cluster correspond to the southern black cross and the northern cross, respectively. }
Figure 1c shows the position-velocity diagram for a strip between the two dashed lines. We find the connecting features at three positions Dec.$\sim$ $\timeform{-18D53'4.4"}$ between the two components at 1525 km s$^{-1}$--1570 km s$^{-1}$ in addition to the other similar features seen at Dec.$\sim$$\timeform{-18D53'5.6"}$ and Dec.$\sim$ $\timeform{-18D53'6.6"}$ (indicated by the arrows in Figure 1c). The position-velocity diagrams for the whole cloud are shown in Appendix 1. Typical spectrum toward integrated intensity peak is shown in Figure 1d. The  average velocity of the blue-shifted cloud and red-shifted cloud is 1500 km s$^{-1}$ and 1600 km s$^{-1}$, respectively. {We identified the velocity components by using moment map, velocity channel maps, and channel maps of declination--velocity diagrams in Figures 6, 7, and 8 by following the method of Paper I. We identified the two velocity ranges to be 1450 km s$^{-1}$--1525 km s$^{-1}$ and 1570 km s$^{-1}$-- 1700 km s$^{-1}$ for the two components.}

Figures 2ab show the distributions of the blue-shifted and red-shifted clouds as indicated {by} the shaded regions of the profile in Figure 1d. Figure 2c shows the distribution of bridge features superposed on the two velocity components. The bridge features are distributed toward overlapping region of blue-shifted cloud and red-shifted cloud where there are many star forming regions as shown by blue crosses in Figure 2c.

Figure 3 shows the distributions of the 1st and 2nd moment toward SSC B1 {in a velocity range of 1450--1700 km s$^{-1}$. }{Figures 3a and 3b show the intensity-weighted velocity (1st moment) and the velocity dispersion (2nd moment), respectively. There are two velocity components at $\sim$1450 km s$^{-1}$ (green) and $\sim$1600 km s$^{-1}$ (red). Figure 3b shows that the velocity dispersion is enhanced to 30--80 km s$^{-1}$ where the red-shifted cloud and the blue-shifted cloud are overlapped as shown in Figure 2c.}

In Figure{s 4a, 4b, and 4c}, average velocity and velocity dispersion of the blue-shifted cloud are 1510 km s$^{-1}$ and 15 km s$^{-1}$, respectively, those of the red-shifted cloud are 1600 km s$^{-1}$ and 30 km s$^{-1}$, respectively{, and those of the bridge are 1536 km s$^{-1}$ and 7 km s$^{-1}$, respectively}. The velocity dispersions of the blue-shifted and red-shifted clouds are enhanced toward western part of cloud where two velocity clouds overlap as shown in Figures {4c and 4d}. 
By using the average velocity and dispersion, we identified the two velocity ranges to be 1450 km s$^{-1}$--1525 km s$^{-1}$ and 1570 km s$^{-1}$--1700 km s$^{-1}$ for the two components.

We calculated the cloud mass of the region enclosed by a contour of 20 \% of the peak integrated intensity by using the CO-to-H$_{2}$ conversion factor $X_{\rm CO}$ = 0.6$\times$10$^{20}$ cm$^{-2}$ (K km s$^{-1}$)$^{-1}$ (Zhu et al. 2003; Kamenetzky et al. 2014). {The column density of molecular hydrogen $N$(H$_{2}$) is calculated by the relationship $N$(H$_{2}$) =$X_{\rm CO}$$\times$$W_{\rm CO}$, where $W_{\rm CO}$ is the integrated intensity of the $^{12}$CO($J$=1--0).} {A} $^{12}$CO($J$=3--2) to $^{12}$CO($J$=1--0) {integrated intensity} ratio {is taken to be} 0.5$\pm$0.1 for SGMC 4/5 (Ueda et al. 2012). 

The molecular masses of the blue-shifted cloud {within an area enclosed by a contour of 219 K km s$^{-1}$ (20\% of the peak integrated intensity) and the red-shifted cloud within an area enclosed by a contour of 173 K km s$^{-1}$ (20\% of the peak integrated intensity) are (4.0--6.0)$\times$10$^7$ $M_{\rm \odot}$ and (1.6--2.4)$\times$10$^7$ $M_{\rm \odot}$, respectively.The smaller value of mass and the larger value of mass correspond to a $^{12}$CO($J$=3--2) to $^{12}$CO($J$=1--0) integrated intensity ratios of 0.4 and 0.6, respectively. } 
We also {calculated} the mass of the bridge component ($V_{\rm LSR}$ = 1525--1570 km s$^{-1}$) {to be (5.2--7.8)$\times$10$^7$ $M_{\odot}$ in the same way with} the blue-shifted and red-shifted clouds. {We define the bridge component as the molecular cloud with $V_{\rm LSR}$ of 1525--1570 km s$^{-1}$ within radius of 200 pc from SSC B1 because the emission in the same velocity range outside of the SGMC 4/5 is not related with cluster formation as shown in Figure 2c.
The bridge component extends to the north, so the mass within a radius of 200 pc from B1 is (1.7--2.5)$\times$10$^7$ $M_{\odot}$. Thus, total mass of the red-shifted, blue-shifted, and bridge clouds around B1 is (6.3--10.9)$\times$10$^7$ $M_{\odot}$. The error of the molecular mass is calculated by uncertainty of the $^{12}$CO($J$=3--2) to $^{12}$CO($J$=1--0) ratio and the RMS noise levels of $^{12}$CO($J$=3--2). In addition, the value of $X_{\rm CO}$ possibly changes {from 0.2$\times$10$^{20}$ to 1.2 $\times$10$^{20}$ cm$^{-2}$ (K km s$^{-1}$)$^{-1}$} because of the difference of the relative abundance of CO to H$_{2}$.}

\begin{figure*}[htbp]
\begin{center}
\includegraphics[width=14cm]{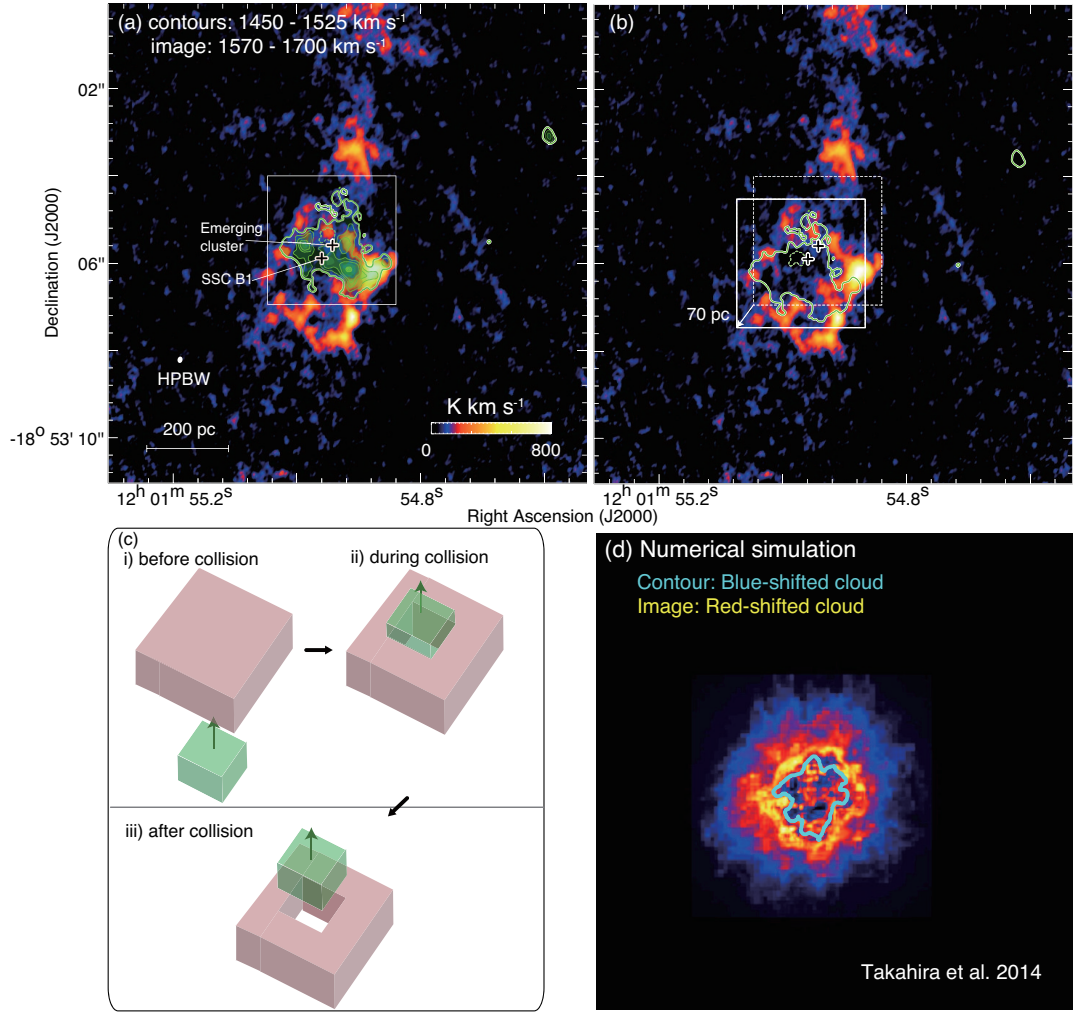}
\end{center}
\caption{(a) CO intensity map of the blue-shifted cloud by contours superposed on the red-shifted cloud. The contour levels and symbols are the same as in Figure 2b. (b) Same as (a), but the contour of blue shifted cloud is displaced. The contour level is 300 K km s$^{-1}$. (c) The projected displacement is 70 pc with a position angle of 126 deg. Right panel shows rectangular solid model clouds before the collision (i), during the collision (ii), and after the collision (iii) for SGMC 4/5. (a) and (b) correspond to (iii) and (ii), respectively. (d) Result of synthetic observation based on the numerical simulation by Takahira et al. (2014). Contour and image indicate blue-shifted cloud and red-shifted cloud, respectively. } 
\label{fig5}
\end{figure*}%

Figure {5} shows comparison of the CO distribution with optical and near infrared image of HST (B-band, V-band, I-band, and H$\alpha$). The image shows two point-like sources {(SSC B1 and {the} emerging cluster classified by Whitmore et al. 2014)} as indicated by two {crosses} in Figure {5}a.
{SSC B1 and the emerging cluster correspond to the southern black cross and the northern cross, respectively.} The two sources correspond to the intensity depression of the blue-shifted cloud (Figure {5}a) and the cavity of the red-shifted cloud to the extended emission ({gray dashed circle in} Figure {5}b). Image of figure {5}cd is H$\alpha$ emission, and correspondence of extended emission and the cavity of red-shifted cloud is remarkable. {A typical radius of the cavity is $\sim$120 pc which is about same size as the blue-shifted cloud.}
The correspondence shows that the two clouds are physically associated with SSC B1 and are the parent clouds of SSC B1. 

{{Whitmore et al. (2014) classified the emerging cluster as a separated object for the first time.} The authors defined the five evolutionary classifications of the SGMCs. Based on the previous study of the GMC evolution model in the Local Group of galaxies (LMC; Fukui et al. 1999, Kawamura et al. 2009, M33; Miura et al. 2012), the authors extended the classification to older clusters with no CO by using radio continuum, optical/near-infrared data in addition to CO.  The emerging cluster is classified as stage 3 (time scale: 0.11 Myr) and gas and dust are removed. Thanks to the higher angular resolution, we found found the intensity depression toward the emerging cluster, which we did not find in Paper I. Thus, it is likely that the two clouds are also associated with the emerging cluster.}

\begin{deluxetable}{ccccccc}
\tablewidth{15cm}
\tablecaption{Physical properties of the {parent cloud of the clusters}}
\label{tab:ssc}
\tablehead{\multicolumn{1}{c}{Object}&$W_{\rm CO}$ [K km s$^{-1}$]\tablenotemark{\dagger}&$L_{\rm H\alpha}$&$R$& {$M_{\rm parent}$} \tablenotemark{\ddagger}& $M_{\rm HII}$ &$M_{cluster}$\\
\multicolumn{1}{c}{\ }&$N_{\rm H_{2}}$ [10$^{22}$ cm$^{-2}$]\tablenotemark{\dagger} & [10$^{38}$ erg s$^{-1}$] &[pc]&[10$^6$ $M_{\odot}$] &[10$^6$ $M_{\odot}$] &[10$^6$ $M_{\odot}$] \\
\multicolumn{1}{c}{(1)}&(2)&(3)&(4)&(5)&(6)&(7)}
\startdata
&3.6&&&&&  \\ 
SSC B1&303&11&19&3.4$^{+3.4}_{-2.3}$&$\sim$1.9&6.8  \\ 
&&&&&  \\ \hline
&2.2&&&&&  \\
Emerging cluster &187&2.4&19&3.8$^{+3.8}_{-2.5}$&$\sim$0.4&---  \\
&&&&&& \\ \hline
Protocluster&15&&&&&  \\
candidate&1266&---&18&2.0$^{+2.0}_{-1.3}$&---&---  \\
ALMA 4&&&&&& \\ \hline
Protocluster&19&&&&&  \\
candidate&1548&---&21&3.4$^{+3.4}_{-2.3}$&---&---  \\
ALMA 5&&&&&& \\ \hline
\enddata

\tablecomments{{Column (1): Object name. Column (2): Column density and integrated intensity of molecular cloud. Column (3):  Luminosity of H$\alpha$ emission obtained with HST (Whitmore et al. 2010). Column (4): The radius of the object. The radius of the protocluster is defined as the size where the integrated intensity decreases to 50\% of the peak integrated intensity. Radii of parent clouds of SSC B1 and the emerging cluster are assumed to be geometric means of radii of the two protoclusters. Column (5): {$M_{\rm parent}$ is molecular mass inside of the radius projected to have been within the radius based on extrapolation from ALMA4 and ALMA5.} The CO cloud masses of SSC B1 and the emerging cluster are given by [(the average $N_{\rm H_{2}}$ of ALMA4 and ALMA5)-(the present $N_{\rm H_{2}}$ toward SSC B1/emerging cluster)]$\times$$\pi$$R$$^2$. Column (6): Ionized gas mass derived by using the relationship between H$\alpha$ luminosity and ionized gas mass $M_{\rm gas \ ionized} = 0.98\times 10^{9} \left(\frac{L_{\rm H\alpha}}{10^{43} \rm erg s^{-1}}\right) {\left(\frac{n_{\rm e}}{100 \rm cm^{-3}}\right)}^{-1}$ (Osterbrock \& Ferland 2006).  $L_{\rm H\alpha}$ is luminosity of H$\alpha$ [erg s$^{-1}$] and $n_{\rm e}$ is assumed to be 100  cm$^{-3}$. Column (7): Total stellar mass of the cluster (Whitmore et al. 2010).}}
\tablenotetext{\dagger}{{Minimum value for SSC B1 and the emerging cluster, and maximum value for the protocluster candidates.}}
\tablenotetext{\ddagger}{{Errors of {$M_{\rm parent}$} are derived from factor three uncertainty of $X_{\rm CO}$ factor.}}
\end{deluxetable}

\subsection{Complementary distribution and bridge features}

Figure {7} shows an overlay of the distributions of the two velocity components from the present analysis, the theoretical simulation result and a schematic of the two clouds. According to the previous works, in two colliding clouds which show complementary distribution, we often find displacement between them (e.g., Fukui et al. 2018; Takahira et al. 2014), which is a general signature for a collision whose velocity makes an angle not close to either 0 deg. or 90 deg. 
{Takahira et al. (2014) made hydrodynamical numerical simulations of two spherical clouds with different radii in {a} head-on collision. In the collision the small cloud creates a cavity having a size of the small cloud in the large cloud. The cavity should have complementary distribution with the small cloud. Fukui et al. (2018) presented synthetic observations of such collision by using the results of Takahira et al. (2014), and showed that the complementary distribution usually has a displacement in the sky because of the projection effect of the collision path which makes a certain non-zero angle to the line of sight.}

Following the algorithm by Fujita et al. (2020a), we moved the red-shifted cloud from the original position over $\pm$ 90 pc with a 1.8 pc (size of pixel) step in the two orthogonal directions of RA and Dec., and calculated the correlation coefficient with the integrated intensity of the blue-shifted cloud for each pixel. 
We calculated Spearman’s correlation coefficient between the integrated intensity of two velocity components, and looked for the position where the correlation coefficient is minimum. 
The method is appropriate when the feedback is not significant and the shapes of the two clouds are well kept after the collision. In Figure {7}a we see the two clouds exhibit a complementary distribution with a possible displacement. \clearpage
Figure {7}b shows the result after the displacement of 70 pc in length and a position angle of $\timeform{126D}$. These parameters give the minimum correlation coefficient of $-$0.7, and their coincidence looks good.
Figure 7c shows a schematic of two clouds before, during and after the collision modified from Figure 14 in {Paper I}.
In Figure 7{d} the two colliding spherical clouds with different radii (Takahira et al. 2014) are overlaid for a case with the collision velocity same with the line of sight.

\section{Discussion}
\subsection{Collision picture toward SSC B1}
{Paper I} suggested that the two clouds are colliding toward SSC B1 based on the complementary distribution and bridge features of the ALMA Cycle 0 data. The present analysis of the Cycle 3 data at four-time higher resolution showed significant details of the complementary distribution and the bridge features of the two clouds, allowing us to elaborate the collision picture. In particular, a displacement of 70 pc with a position angle of $\timeform{126D}$ was derived by the fitting procedure between the small cloud and the cavity of the large cloud (Fujita et al. 2020; see also Fukui et al. 2018a).

The new collision picture above helps to estimate better collision parameters. The collision time scale is calculated by a ratio of the travelled distance $\sim$70 pc divided by the collision velocity $\sim$100 km s$^{-1}$ to be 0.7 Myr. The time scale is valid if the collision angle is 45 deg. If the collision direction makes an angle of 30 deg. or 60 deg. to the line of sight, the time scale can vary from 0.4 to 1.4 Myrs corrected for the projection. This value is consistent with the estimated cluster age, 1--3.5 Myrs (Gilbert \& Graham 2007; Whitmore et al. 2010).

According to the picture the blue-shifted small cloud collided close to the center of the red-shifted large cloud, where the small cloud moved from the south to the north and fully passed through the large cloud. As a result, {the collision could have created the cavity in the large cloud,} and compressed the interface layer between the two clouds (a schematic diagram in Figure {7}). Part of the remnant of the interface layer is seen as the bridge stellar objects (Figure 2c), while the rest of the gas has been ionized. The near infrared bright and compact features in the both clouds correspond to the intensity depressions toward the stellar objects and the extended feature associated. This suggests that part of the two clouds is converted to the stars of the cluster in the collision.

\subsection{{Mass budget and star formation efficiency}}
{We estimate the mass budget and the SFE in the two colliding clouds by considering the gas mass, stellar mass, ionized gas mass, and extrapolating the gas mass of the parental cloud before formation.\\
\underline{{Mass budget}}\\{We attempt to estimate the star formation efficiency in the colliding clouds.} Figure 2 shows that the two clouds are clearly separated with a displacement of 70 pc. We adopt an $X_{\rm CO}$ factor of 0.6$\times$10$^{20}$ cm$^{-2}$ (K km s$^{-1}$)$^{-1}$ which has {large uncertainty} (Zhu et al. 2003; Kamenetzky et al. 2014), and the uncertainty in the molecular mass is mainly due to $X_{\rm CO}$. {}
{For the $X_{\rm CO}$} red-shifted cloud and the blue-shifted cloud have 2.0$^{+0.4}_{-0.4}$$\times$ 10$^7$ $M_{\rm \odot}$ and 5.0$^{+1.0}_{-1.0}$$\times$10$^7$ $M_{\rm \odot}$, respectively{, where only the error bars in an intensity ratio of the $J$=3--2 and 1--0 transitions are considered}. Figure 5 shows that the formation of the two clusters, SSCB1 and the emerging cluster, is taking place in the blue-shifted cloud as evidenced by the {two} depressions of the molecular gas accompanying the H$\alpha$ emission. The blue-shifted cloud has {two additional} molecular peaks {outside the two clusters,} which are suggested by Whitmore et al. (2014) to be {two more} candidate protoclusters as indicated in Figure 5{, whereas they have no stellar emission}. {The gas masses of the protocluster candidates ALMA 4 and ALMA 5 are calculated to be 2.0 $\times$10$^6$ $M_{\odot}$ and 3.4 $\times$ 10$^6$ $M_{\odot}$ above an integrated intensity level of 633 K km s$^{-1}$ and 774 K km s$^{-1}$ which correspond to 50\% of the peak integrated intensity, respectively. The sizes of ALMA 4 and ALMA 5 are 18 pc and 21 pc in radius, respectively.}}


{In order to estimate the molecular mass which formed the two clusters SSC B1 and the emerging cluster, we interpolate the mass which existed in the two cavities prior to the collision. It is probable that the molecular gas is depleted by the cluster formation and the resultant ionization.} {We assume radius $R$ of parent cloud of SSC B1 and the emerging cluster are the geometric mean of $R$ of the two protocluster candidates. We  estimate $R$ of the depressions to be $\sim$19 pc for SSC B1 and the emerging cluster. We assumed that the initial $N_{\rm H_{2}}$ toward SSCB1 prior to the cluster formation is equal to an average of the peak $N_{\rm H_{2}}$ toward the two protocluster candidates ALMA4 and ALMA5. Then, the parent cloud mass {$M_{\rm parent}$} is given by [(the average $N_{\rm H_{2}}$ of ALMA4 and ALMA5)$-$(the present $N_{\rm H_{2}}$ toward SSC B1)]$\times$$\pi$$R$$^2$. We estimate the molecular mass inside the radius to be 3.4$^{+3.4}_{-2.3}$$\times$10$^6$ $M_{\odot}$. A similar method is applied to the emerging cluster and we estimate a radius of the parent cloud to be 19 pc for the emerging cluster, and find that the mass in the radius is 3.8$^{+3.8}_{-2.5}$$\times$10$^6$ $M_{\odot}$. The relevant figures are listed in Table 1. }


{Further, we use the H$\alpha$ emission to calculate the ionized gas mass by adopting the following relationship $M_{\rm gas \ ionized} = 0.98\times 10^{9} \left(\frac{L_{\rm H\alpha}}{10^{43} \rm erg s^{-1}}\right) {\left(\frac{n_{\rm e}}{100 \rm cm^{-3}}\right)}^{-1}$, where $L_{\rm H\alpha}$ is luminosity of H$\alpha$ in units of erg s$^{-1}$ and $n_{\rm e}$ is averaged electron density in units of cm$^{-3}$ (Osterbrock \& Ferland 2006). We assume $n_{\rm e}$ to be 100 cm$^{-3}$ from the size-density relation of extragalactic H{\sc ii} regions \citep{2009A&A...507.1327H}. Values of $L_{H\alpha}$ of SSC B1 and the emerging cluster are estimated to be 1.1$\times$10$^{39}$ erg s$^{-1}$, 2.4$\times$10$^{38}$ erg s$^{-1}$, respectively, and we calculate the ionized gas mass of SSC B1 and the emerging cluster are 2$\times$10$^6$ $M_{\odot}$ and 0.4$\times$10$^6$ $M_{\odot}$, respectively. These are an order of magnitude smaller than the molecular mass at most.}\\ \underline{{Star formation efficiency}}\\ 
{The cluster mass of SSC B1 is estimated to be 6.8$\times$10$^6$ $M_{\odot}$ (Whitmore et al. 2010) and 4.2$\times$10$^6$ $M_{\odot}$ (Gilbert \& Graham 2007). The former is corrected for the extinction and we adopt it in following. The peak luminosity of H$\alpha$ of SSC B1 is greater by an order of magnitude than that of the emerging cluster. This suggests that the emerging cluster is young and has not yet ionized its surrounding gas. The cluster mass of the emerging cluster therefore must be crude at best. Table 1 lists $W_{\rm CO}$ and the extrapolated parent cloud mass of the two clusters and two protoclusters. }\\
{Except for SSC B1, there is no reliable estimate of the cluster mass, and we focus on SSC B1. By using the cluster mass and the extrapolated molecular mass, we estimate the star formation efficiency (SFE) to be a ratio of the cluster mass divided by {the total mass of extrapolated molecular mass and ionized gas mass}, 6.8$\times$10$^6$ $M_{\odot}$/[(3.0--8.7)$\times$10$^6$ $M_{\odot}$]$\sim$0.78, if we assume an $X_{\rm CO}$ {value ranges from 0.2$\times$10$^{20}$ to 1.2$\times$10$^{20}$ depending on the difference of the relative abundance of CO to H$_{2}$ (Zhu et al. 2003). We assume that the relative abundance of CO to H$_{2}$ possibly changes from 5$\times$10$^{-5}$ to 2.7 $\times$10$^{-4}$ by observations of GMCs in the star forming regions of the Milky Way and the theoretical model (e.g., Blake et al. 1987; Farquhar, Milar, \& Herbst 1994).}
Obviously, SFE greater than 1 does not make sense. {One interpretation of this extremely high SFE is that the calculation of the extrapolated parental cloud mass is too small. This suggests that the column density of the parental cloud was  larger than the column density of these other proto-clusters.}
We infer that the local cluster formation within 20 pc in radius is taking place at very high SFE more than 0.78. Over the whole collision process, the total molecular mass of the blue-shifted cloud is $\sim$2$\times$10$^8$ $M_{\odot}$ at 200--300 pc scale, and SFE for SSC B1 only becomes $\sim$3\%. If we include the mass of the emerging cluster and the two protocluster candidates, the total SFE becomes $\sim$12\% under an assumption of their equal cluster mass. So, the cloud-cloud collision may not be an extremely efficient process in converting the cloud mass into clusters in the SGMC 4--5 region.}

\subsection{{SSC formation and physical condition of around environment}}

{It is suggested that the high-pressure environment is a requirement for the SSC formation (Elmegreen \& Efremov 1989; Elmegreen \& Efremov 1997). Johnson et al. (2015) and Finn et al. (2019) found the high-pressure environment in the overlap region. Furthermore, {Paper I} revealed a positive correlation between the compressed gas pressure generated by collision and total stellar mass of cluster. }
The typical pressure of the colliding clouds is estimated to be {(1.7--8.0)}$\times$10$^{8}$ K cm$^{-3}$ by using the relationship, $P_{\rm e}= \frac{3\Pi \it M v^{2}}{4 \pi R^{3}} = \rho_{e} v^{2}$, which is consistent with the relationship derived by {Paper I}.  $M$ is the cloud mass, $v$ is the colliding velocity (difference of peak velocity of colliding clouds), $R$ is the radius of cloud, and $\rho_{e}$ is the density at the cloud edge. $\Pi$ is defined by $\rho_{e}$=$\Pi$$\rho$, where $\rho$ is the mean density in the cloud. We adopt $\Pi$ = 0.5 (Jonson et al. 2015; Finn et al. 2019). {The error of the external pressure mainly consists of the uncertainty of the mass estimation and the projection effect on the collision velocity. If we assume an angle of the collision path to the line of sight $\theta$ is 30--60 deg., the collision velocity could be ranged from $\sim$115 km s$^{-1}$ to $\sim$200 km s$^{-1}$. 
For $\theta$= 90 deg., we do not see the two velocity components and for $\theta$ = 0 deg., we do not see the displacement between the two components as shown in Figure 6.}

Feedback is believed to be important in high-mass star formation. The present results reveal some novel aspects of the feedback. Figure {4} shows the distribution of {1st moment} and {2nd moment} {of the blue-shifted, red-shifted, and bridge clouds}. The stellar objects in the cluster is plotted by crosses and show no variation of the average velocity or the velocity dispersion is seen, suggesting that the stellar feedback is not affecting appreciably the gas dynamics. {Figure 3 also shows the spatial distribution of {1st moment} and {2nd moment} at the all velocity range ($V_{\rm LSR}$). The velocity dispersion is raised to $\sim$15 km s$^{-1}$ toward the stellar objects, which is possibly induced by stellar feedback at several tens pc scale. On the other hand, the maximum velocity dispersion is $\sim$80 km s$^{-1}$ in the area where the red-shifted cloud and the blue-shifted cloud are overlapped. From these results, it is natural to understand the velocity structures of the whole SGMC 4/5 by the motion of the red-shifted and blue-shifted clouds, not by feedback.} 

We suggest that the velocity distribution of the clouds is more influenced by the collisional interaction than the feedback as indicated by the enhancement of {2nd moment} having a spatial gradient from the east to west which has no correlation with the stellar objects. The area of the cluster formation is strongly {ionized} by the forming stars in the cluster in the both clouds as depicted in Figure {5}. The present high resolution shows the cavity is fully enclosed by the large cloud (Figure {7}d).

\section{Conclusion}
The Antennae galaxies is the most outstanding major merger closest to the Milky Way. We analyzed the ALMA Cycle3 dataset of the $^{12}$CO($J$=3--2) emission toward the SSC B1 and derived the molecular distribution at 10 pc resolution. The main conclusions of the present study is summarized as follows;

\begin{enumerate}
\item The present results confirm that the molecular gas in SSC B1 has two velocity peaks at 1500 km s$^{-1}$ and 1600 km s$^{-1}$ (Paper I). The spatial distributions of the clouds resolved at a resolution of 10 pc shows complementary distribution with a displacement of 70 pc which was determined more accurately than in Paper I. The two clouds are connected with bridge features in the intermediate velocity range. The complementary distribution well matches the numerical simulation by Takahira et al. (2014).

\item {The star formation efficiency (SFE) of SSC B1 is estimated to be larger than $\sim$0.78 by the total mass including the extrapolated molecular mass and ionized gas at 20 pc. SFE is very high at 20 pc scale. On the other hand, over the whole collision process, the total mass of the blue-shifted cloud associated with the cluster formation is $\sim$2$\times$10$^8$ $M_{\odot}$ at 200-300 pc scale, and SFE is only $\sim$3\%. If we include the total stellar mass of the emerging cluster and two protocluster candidates, the SFE becomes $\sim$12\% for assumption of their equal cluster mass. Thus, the cloud-cloud collision may not be an extreme efficient process in the SGMC 4--5 region. }


\item The velocity distribution in the two CO clouds are derived as 1st {moment} and 2nd {moment}, and are compared with compact near infrared sources and the H$\alpha$ image. The result shows that the gas motion has no sign of perturbed velocity field by the stellar sources. We therefore argue that significant kinematic feedback by the newly formed stars is not important except for the ionization which probably evacuated the molecular gas. {We found two cavities formed by ionization of clusters and estimated ionized gas mass was $\sim$(0.4--1.9)$\times$10$^6$ $M_{\rm \odot}$.} This suggests that the cloud kinematics is dominated by the galactic tidal interaction.

\end{enumerate}

\begin{ack}
{This paper makes use of the following ALMA data: ADS/ JAO.ALMA \#2011.0.00876, and \#2015.1.00038.S . ALMA is a partnership of the ESO, NSF, NINS, NRC, MOST, and ASIAA. The Joint ALMA Observatory is operated by the ESO, AUI/NRAO, and NAOJ. {This paper is also based on observations made with the NASA/ESA Hubble Space Telescope, obtained from the data archive at the Space Telescope Science Institute. STScI is operated by the Association of Universities for Research in Astronomy, Inc. under NASA contract NAS 5-26555.} 
K. Tsuge was supported by the ALMA Japan Research Grant of NAOJ ALMA Project, NAOJ-ALMA-232. This study was financially supported by JSPS KAKENHI ({Grant Numbers 15H05694 and 18K13582}). {We would like to thank the referees for their constructive inputs.} }
\end{ack}

{}

\appendix
\section{CO channel maps of position diagrams}
We show the 21 declination-velocity diagrams of the $^{12}$CO($J$=3--2). The integration range is 0$\farcs$5 ($\sim$54 pc), and the integration range is shifted from east to west in 0$\farcs$5 step.

\begin{figure*}[htbp]
\begin{center}
\includegraphics[width=16cm]{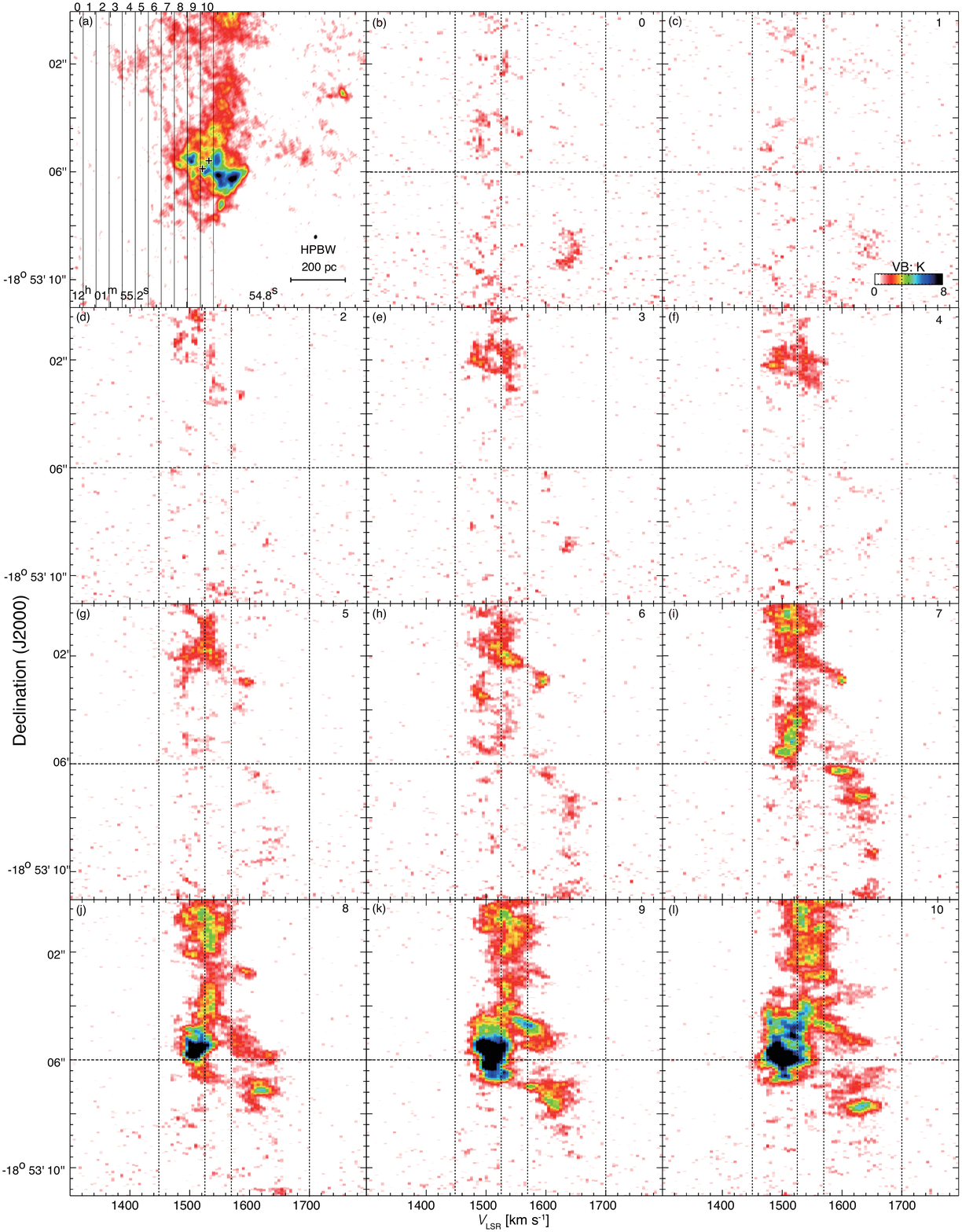}
\end{center}
\caption{Channel maps of declination-velocity diagrams. (a) Total integrated intensity map of SGMC 4/5. The integration velocity range is the same as in Figure 1b. Dashed vertical lines indicate the integration ranges of declination-velocity diagrams in R.A.. (b)--(l) Declination-velocity diagrams of $^{12}$CO ($J$=3--2). The upper right {number} denotes the integration range in (a).} 
\label{figa2}
\end{figure*}%

\begin{figure*}[htbp]
\begin{center}
\includegraphics[width=16cm]{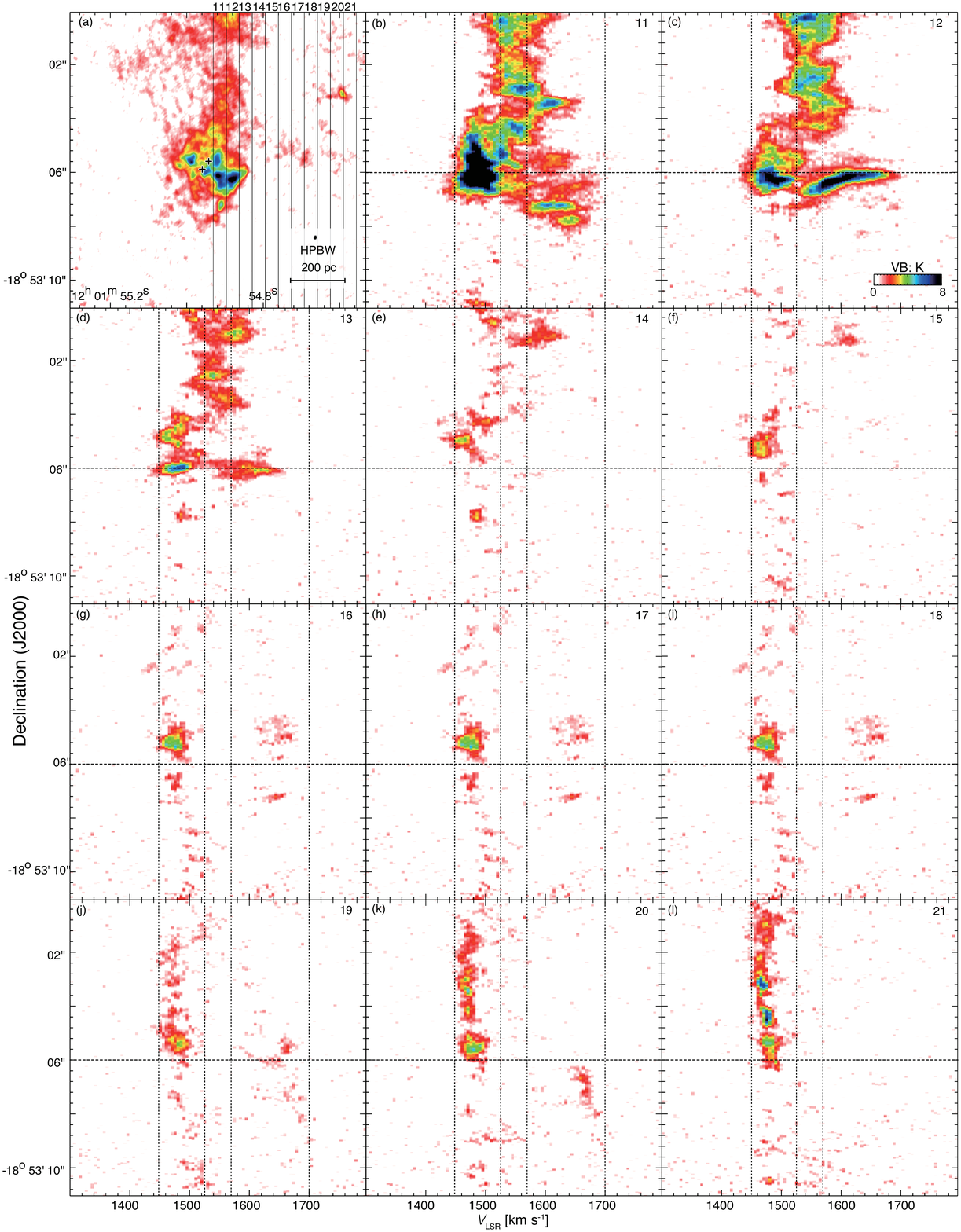}
\end{center}
\caption{Channel maps of declination-velocity diagrams. (a) Total integrated intensity map of SGMC 4/5. The integration velocity range is the same as in Figure 1b. Dashed vertical lines indicate the integration ranges of declination-velocity diagrams in R.A.. (b)--(l) Declination-velocity diagrams of $^{12}$CO ($J$=3--2). The upper right {number} denotes the integration range in (a).} 
\label{figa3}
\end{figure*}%

\section{Velocity channel maps}
We show the velocity channel maps of the $^{12}$CO($J$=3--2) of SGMC 4/5 in Figure 8. The velocity range is between 1400 km s$^{-1}$ and 1700 km s$^{-1}$.

\begin{figure*}[htbp]
\begin{center}
\includegraphics[width=16cm]{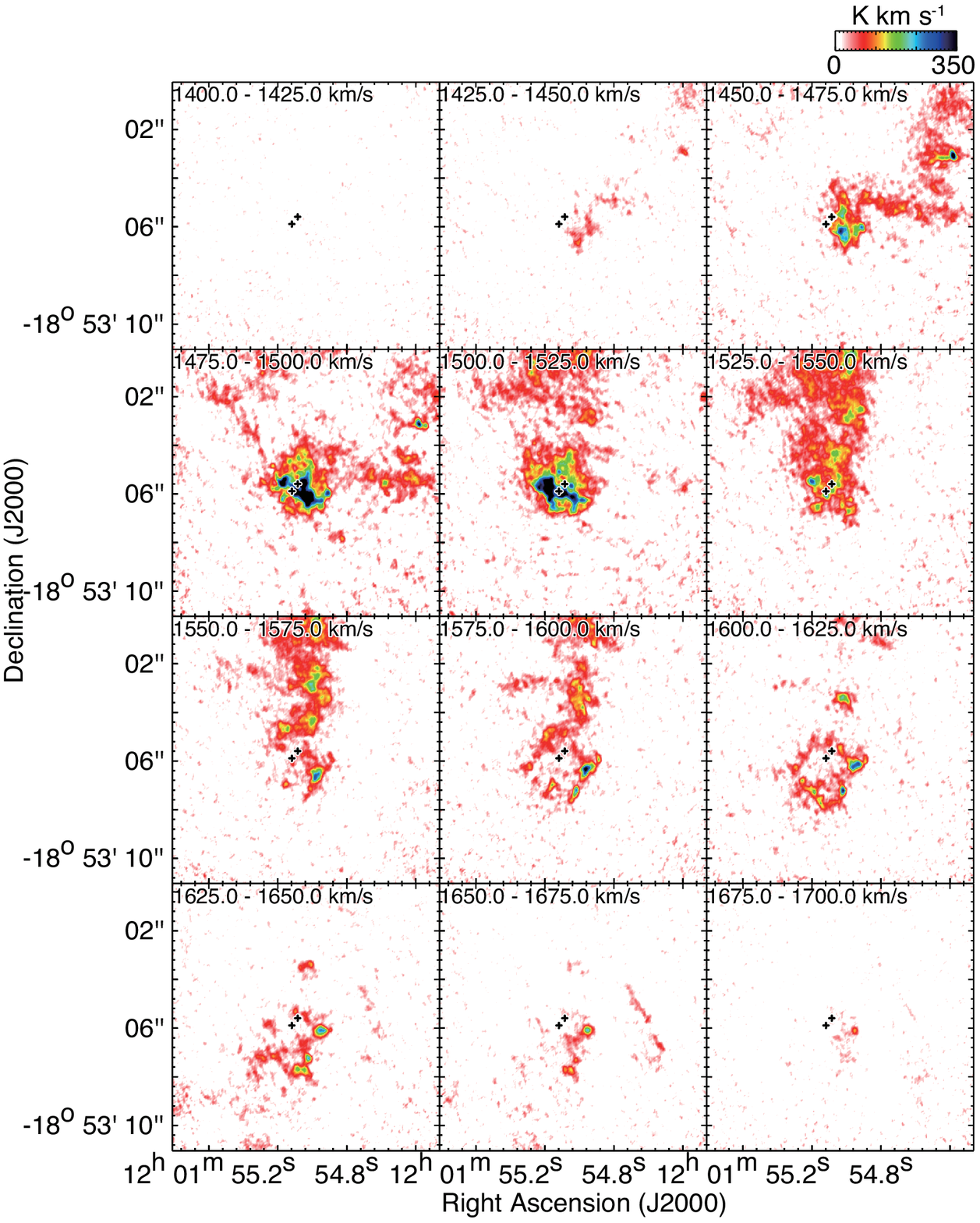}
\end{center}
\caption{Velocity channel maps of $^{12}$CO(3--2) toward the SGMC 4/5 with velocity step pf 25 km s$^{-1}$. {The symbols are the same as in Figure 1.} } 
\label{figa1}
\end{figure*}%

\end{document}